\documentclass[12pt,a4paper]{article}
\usepackage{amsmath}
\usepackage{bm}
\usepackage{amsfonts}
\usepackage{graphicx}
\usepackage[normal]{caption2}
\usepackage{subfigure}
\usepackage{rotating}
\usepackage{citesort}
\usepackage{lscape}
\usepackage{section}
\begin{document}
\renewcommand\arraystretch{1.1}
\setlength{\abovecaptionskip}{0.1cm}
\setlength{\belowcaptionskip}{0.5cm}
\title { The study of participant-spectator matter and collision
dynamics in heavy-ion collisions}
\author {Aman D. Sood and  Rajeev K. Puri\\
\it Department of Physics, Panjab University, Chandigarh -160 014,
India.\\}
\maketitle
\begin{abstract}
We aim to study the participant-spectator matter over a wide range
of energies of vanishing flow and masses. For this, we employed
different model parameters at central and semi-central colliding
geometries. Remarkably, a nearly mass independent nature of the
participant matter was obtained at the energy of vanishing flow.
This makes it a very strong alternative candidate to study the
energy of vanishing flow. We also show that the participant matter
can also act as an indicator to study the degree of
thermalization. The degree of thermalization reached in central
collisions is nearly the same for different colliding nuclei at
the energy of vanishing flow.
\end{abstract}
 Electronic address:~rkpuri@pu.ac.in
\newpage
\section{Introduction}
One of the main goals of heavy-ion collisions at intermediate
energy is to study the nuclear properties of hot and dense
environment. This is influenced by the nuclear matter equations of
state as well as by the in-medium nucleon-nucleon cross
sections\cite{stoc86,aich91,grei88}. In addition, the incident
energy, the mass of colliding system as well as impact parameter
also has a strong influence\cite{hart98,puri94,gupt88,khoa92548}.
The nuclear interactions are attractive at low incident energies
allowing the scattering of nucleons into backward hemi-sphere.
These interactions, however turn repulsive, once incident energy
is high enough. In this energy variation, these interactions
counterbalance each other, at a particular point. This leads to no
net preference to the nuclear transverse in-plane flow, therefore,
net nuclear flow disappears.
\cite{ogil90,krof92,mage0061,mage0062,cuss02,west93,sull90,ange97,krof91,he96,buta95,pak97,zhan90,li93,mota92,xu92,zhou94a,lehm96,soff95,kuma98,sood04}
This energy of vanishing flow (EVF) has always been of great
interest. It gives a possibility to extract the nuclear matter
equation of state and strength of in-medium nucleon-nucleon cross
section.
\cite{ogil90,krof92,mage0061,mage0062,west93,sull90,mota92}

Recently, we presented, for the first time, a complete study of
the energy of vanishing flow over the entire periodic
table\cite{sood04}. There, a mass power law was obtained and a
very close agreement with the experimental observations was also
found\cite{sood04}. In another communication, we also studied the
other phenomena of heavy-ion reactions at the energy of vanishing
flow. For the first time, we reported that the
participant-spectator matter at EVF is quite insensitive to the
mass of the colliding system. It, therefore, can act as a
barometer for the study of the energy of vanishing
flow\cite{sood04a}. Here, we plan to extend the study of vanishing
flow for other equations of state as well as for the momentum
dependent interactions. We plan to look whether the above
participant-spectator matter demonstration still holds good or
not. This will also give us a clue whether the
participant-spectator matter at the point of vanishing flow is
indeed an indicator of the counterbalancing of the attractive and
repulsive interactions or is only an accidental coincident.
Section 2 describes the model in brief. Section 3 explains the
results and discussion and section 4 summarizes the results.

\section{The model}

The present study is conducted within the framework of quantum
molecular dynamics (QMD)
model\cite{aich91,grei88,hart98,puri94,khoa92548,lehm96,soff95,kuma98,sood04,sood04a}.
In the quantum molecular dynamics model, each nucleon propagates
under the influence of mutual two and three-body interactions. The
propagation is governed by the classical equations of motion:
\begin{equation}
\dot{{\bf r}}_i~=~\frac{\partial H}{\partial{\bf p}_i}; ~\dot{{\bf
p}}_i~=~-\frac{\partial H}{\partial{\bf r}_i},
\end{equation}
where H stands for the Hamiltonian which is given by:
\begin{equation}
H = \sum_i^{A} {\frac{{\bf p}_i^2}{2m_i}} + \sum_i^{A}
({V_i^{Skyrme} + V_i^{Yuk} + V_i^{Coul} + V_i^{mdi}}).
\end{equation}
The $V_{i}^{Skyrme}$, $V_{i}^{Yuk}$, $V_{i}^{Coul}$, and
$V_i^{mdi}$ are, respectively, the Skyrme, Yukawa, Coulomb, and
momentum dependent potentials (MDI). The momentum dependent
interactions are obtained by parameterizing the momentum
dependence of the real part of the optical potential. The final
form of the potential reads as \cite{aich91}
\begin{equation}
U^{mdi}\approx t_{4}ln^{2}[t_{5}({\bf p_{1}}-{\bf
p_{2}})^{2}+1]\delta({\bf r_{1}}-{\bf r_{2}}).
\end{equation}
Here $t_{4}$ = 1.57 MeV and $t_{5}$ = $5\times 10^{-4} MeV^{-2}$.
A parameterized form of the local plus MDI potential is given by
\begin{equation}
U=\alpha \left({\frac {\rho}{\rho_{0}}}\right) + \beta
\left({\frac {\rho}{\rho_{0}}}\right)^{\gamma}+ \delta
ln^{2}[\epsilon(\rho/\rho_{0})^{2/3}+1]\rho/\rho_{0}.
\end{equation}
The parameters $\alpha$, $\beta$, $\gamma$, $\delta$, and
$\epsilon$ are listed in Ref \cite{aich91}. We shall use both the
soft and hard equations of state. As explained in Ref
\cite{sood04}. and others,\cite{sull90,xu92,zhou94a,kuma98} an
isotropic and constant nucleon-nucleon cross section between 40
and 55 mb is also employed.

\section{Results and discussion}

As stated in the introduction, the energy of vanishing flow is of
great interest. Till now, one has measured and studied
theoretically the energy of vanishing flow in $^{12}$C+$^{12}$C
($b/b_{max}=0.4$), $^{20}$Ne+$^{27}$Al ($b/b_{max}=0.4$),
$^{36}$Ar+$^{27}$Al ($b=2$ fm), $^{40}$Ar+$^{27}$Al ($b=1.6$ fm),
$^{40}$Ar+$^{45}$Sc ($b/b_{max}=0.4$), $^{40}$Ar+$^{51}$V
($b/b_{max}=0.3$), $^{40}$Ar+$^{58}$Ni ($b$ = 0-3 fm),
$^{64}$Zn+$^{48}$Ti ($b=2$ fm), $^{58}$Ni+$^{58}$Ni
($b/b_{max}=0.28$), $^{64}$Zn+$^{58}$Ni ($b=2$ fm),
$^{86}$Kr+$^{93}$Nb ($b/b_{max}=0.4$), $^{93}$Nb+$^{93}$Nb
($b/b_{max}=0.3$), $^{129}$Xe+$^{Nat}$Sn ($b$ = 0-3 fm),
$^{139}$La+$^{139}$La ($b/b_{max}=0.3$), and $^{197}$Au+$^{197}$Au
($b=2.5$
fm)\cite{ogil90,krof92,mage0061,mage0062,cuss02,west93,sull90,ange97,krof91,he96,buta95,pak97,zhan90,li93,mota92,xu92,zhou94a,lehm96,soff95,kuma98,sood04}.
Note that these studies have a couple of limitations in general.

(a) The choice of impact parameter in the above reactions is not
fixed. Rather, it varies between 0.1 to 0.4 of the maximum
permissible value. It has been reported by many
authors\cite{mage0062,sull90,ange97,he96,buta95,pak97,soff95,kuma98,blat91}
that the impact parameter variation could have drastic effects on
the sensitive quantities like the collective flow. Therefore, this
huge variation in the impact parameter should be minimized.

(b) The masses of the colliding nuclei (chosen in the experimental
studies) are  asymmetric in nature in general. They have large
variation in the mass of projectile and target. It is well known
that the asymmetric nuclei do not follow the dynamics of the
symmetric nuclei at the first place. Secondly, the asymmetry of
the reaction has also a significant influence on the collective
flow and on the reaction dynamics \cite{pan93}. In order to have a
meaningful and systematic study, we keep both these factors under
control. We take symmetric colliding nuclei and keep impact
parameter fixed. Therefore, here, we simulated the reactions of
$^{12}$C+$^{12}$C, $^{20}$Ne+$^{20}$Ne, $^{40}$Ca+$^{40}$Ca,
$^{58}$Ni+$^{58}$Ni, $^{93}$Nb+$^{93}$Nb, $^{131}$Xe+$^{131}$Xe,
$^{197}$Au+$^{197}$Au, and $^{238}$U+$^{238}$U. We restricted to
different scaled colliding geometries (i.e., in terms of the
combined radius of target and projectile). The static and momentum
dependent interactions along with nucleon-nucleon cross sections
of 40 and 55 mb are used. The hard (soft) equation of state has
been dubbed as Hard (Soft). Whereas hard equation of state with
MDI has been labeled as HMD. The magnitude of the cross section
appears as a superscript to these abbreviations. The above
mentioned simulations were carried out at small energy steps
between 30 MeV/nucleon and 250 MeV/nucleon. A straight line
interpolation was used to extract the value of the energy of
vanishing flow. The extraction and discussion has been reported
elsewhere \cite{sood04}.
\begin{figure}[!t] \centering
 \vskip 0.5cm
\includegraphics[angle=0,width=10cm]{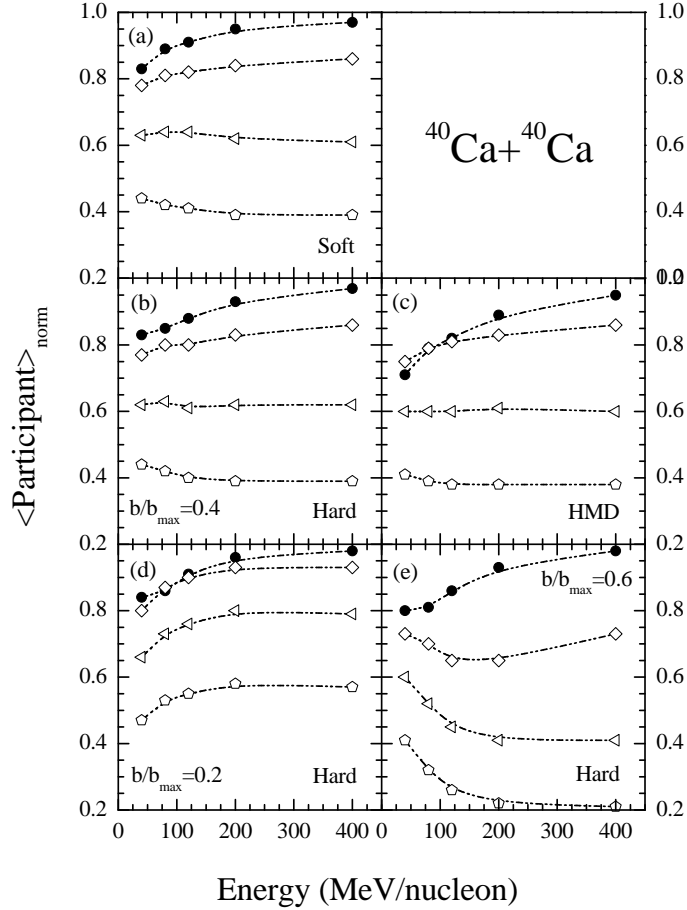}
\vskip -0.2 cm \caption {Top panel: The participant matter as a
function of incident energy for the reaction of
$^{40}$Ca+$^{40}$Ca. Parts \emph{a}, \emph{b}, \emph{c}, and
\emph{d} are for Soft$^{55}$, Hard$^{40}$, Hard$^{55}$, and
HMD$^{55}$, respectively, using $b/b_{max} = 0.4$. Parts \emph{e}
and \emph{f}are for $b/b_{max} = 0.2$ and $b/b_{max} = 0.6$,
respectively, using a hard equation state with $\sigma=55$ mb. The
solid symbols represent the participant matter defined in terms of
nucleon-nucleon collisions. The open symbols are for the
participant matter defined in terms of different rapidity
distribution cuts (see text for details). Lines are only to guide
the eye.}\label{fig1}
\end{figure}
Once the energy of vanishing flow is known, the
participant-spectator matter is extracted with two different
procedures as explained in Ref \cite{sood04a}. (a) All nucleons
having experienced at least one collision are counted as {\it
participant matter} (labeled as PM-C). The remaining matter is
labeled as {\it spectator matter} (labeled as SP-C). The nucleons
with more than one collision are labeled as {\it super-participant
matter} (labeled as SPM-C). These definitions give us a
possibility to analyze the reaction in terms of the
participant-spectator fireball model. These definitions, however,
are more  of a theoretical interest since the matter defined in
these zones cannot be measured. (b) Alternately, we define the
participant and spectator matter in terms of  the rapidity
distribution. The rapidity of \emph{i}th particle is defined as
\begin{equation}
Y(i)= \frac{1}{2}\ln\frac{{\bf{E}}(i)+{\bf{p}}_{z}(i)}
{{\bf{E}}(i)-{\bf{p}}_{z}(i)},
\end{equation}
where ${{\bf E}(i)} $ and $ {\bf p}_{z}(i)$, are, respectively,
the total energy and longitudinal momentum of $i$th particle. We
shall rather use a reduced rapidity
$Y_{red}(i)=Y_{c.m.}(i)/Y_{beam}$. Here different cuts in the
rapidity distributions can be imposed to define the different
participant matter. This method is often used in literature for
experimental analysis \cite{fopi}. We shall define normalized
participant matter by imposing three different cuts: (i) all
nucleons with $ -1.0 \leq Y_{red}(i) \leq +1.0$ (labeled as
PM-R1), (ii)  $ -0.75 \leq Y_{red}(i) \leq +0.75$ ( marked as
PM-R2) and (iii) $ -0.5 \leq Y_{red}(i) \leq + 0.5$ (marked as
PM-R3). These three different definitions  give us possibility to
examine the participant matter at EVF. This can also be verified
experimentally.

First of all, we look for the participant-spectator matter in the
incident energy plane. In Fig. \ref{fig1}, we display the
participant matter as a function of the incident energy for the
reaction of $^{40}$Ca+$^{40}$Ca. The solid symbols represent PM-C
whereas open symbols are for different rapidity distribution cuts.
The open diamonds, open left triangles, and open pentagons
represent, respectively, PM-R1, PM-R2, and PM-R3. The different
sub-figures are for the different sets of model parameters.

First of all, PM-C increases linearly with increase in the
incident energy in all the cases. The increase in the frequency of
nucleon-nucleon collisions is responsible for this linear
response. At incident energies below 200 MeV/nucleon, PM-C depends
significantly on the momentum dependence of the interactions since
at low incident energies, most of the collisions are blocked,
resulting in less significant dependence of PM-C on individual
collisions. Interestingly, PM-C is insensitive towards the impact
parameter at all the incident energies. On the other hand, PM-R1,
PM-R2, and PM-R3 are insensitive to the model parameters. However,
they depend strongly on the impact parameter. The rapidity is also
an indicator of the degree of thermalization reached in a
reaction. The thermalization has been known to be insensitive
towards the EOS and MDI. Whereas it depends strongly on the impact
parameter \cite{khoa92548}. Interestingly, the energy dependence
of the participant matter (defined in terms of different rapidity
distribution cuts) varies with the impact parameters. At central
colliding geometries ($b/b_{max}=0.2$), PM-R1, PM-R2, and PM-R3
increase sharply up to mild energies in agreement with Ref
\cite{khoa92548}. At semi-central geometries ($b/b_{max}=0.4$),
PM-R1 increases, whereas PM-R2 remains constant. On the other
hand, PM-R3 decreases by small amount with increase in the
incident energy. However, at semiperipheral geometries
($b/b_{max}=0.6$), all three participant matters (i.e., PM-R1,
PM-R2, and PM-R3) decrease with increase in the incident energy
which may be due to the fact that the dominance of mean field at
low incident energies and higher impact parameter causes the
thermalization in a reaction. It has been reported in Ref
\cite{sood04a}. that the nature of participant-spectator matter is
not altered by the mass of the colliding nuclei. Of course, the
incident energy of the projectile has a significant role to play.
\begin{figure}[!t] \centering
 \vskip 0.5cm
\includegraphics[angle=0,width=10cm]{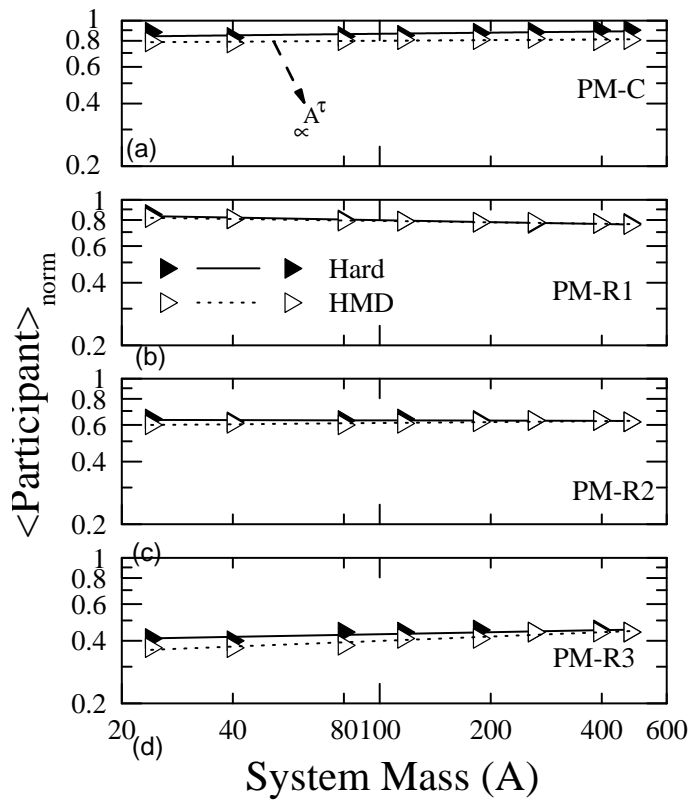}
\vskip -3 cm \caption {Top panel: The participant matter as a
function of system size at their corresponding energy of vanishing
flow. Here we use $b/b_{max} = 0.4$ and different model
parameters. Parts \emph{a} and \emph{b} are for participant matter
defined in terms of nucleon-nucleon collisions. Parts \emph{c},
\emph{d}, and \emph{e} are for participant matter defined in terms
of different rapidity distribution cuts. Lines are power law fit
($\propto A^{\tau}$).}\label{fig2}
\end{figure}

Let us now discuss the participant-spectator matter at the energy
of vanishing flow. As reported in
Ref.,\cite{mage0061,mage0062,west93,mota92,sood04} the energy of
vanishing flow exhibits ($A^{-\tau}$) power law mass dependence.
Naturally, the lighter colliding nuclei have higher balance energy
whereas heavier colliding nuclei have smaller balance energy. This
is true for every equation of state with or without momentum
dependence as well as for different cross sections. It was argued
that due to small number of nucleon-nucleon collisions in lighter
colliding nuclei, one needs higher incident energies to reach the
balance point. In our earlier report,\cite{sood04a}
participant-spectator matter has been reported to be almost
constant at EVF for entire periodic table in central collisions.
This happens because of the fact that at the balance energy, the
attractive and repulsive forces counterbalance each other. This
leads to the same amount of participant-spectator matter. We here
extended this concept for different colliding geometries and also
for different model ingredients to see whether the above concept
is still valid or not. It is worth mentioning that the variation
in EVF for a particular colliding mass using different model
ingredients can be as large as 75 MeV/nucleon.

In Fig. \ref{fig2}, we display the participant matter as a
function of combined mass of the system at their corresponding
EVF. In Fig. \ref{fig2}a and \ref{fig2}b, we display,
respectively, the PM-C and SPM-C. In Fig. \ref{fig2}c,
\ref{fig2}d, and \ref{fig2}e, we display (the participant matter
defined in terms of different rapidity distribution cuts) PM-R1,
PM-R2, and PM-R3, respectively. The solid (open) squares are for
Hard$^{40}$ (HMD$^{40}$). The solid (open) right triangles are for
Hard$^{55}$ (HMD$^{55}$). The lines are the power law fit of the
form $cA^{\tau}$. The solid (dotted) lines are for the hard (HMD)
equation of state. All the reactions are at the semi-central
geometry of $b/b_{max}=0.4$. From Fig. \ref{fig2}a and
\ref{fig2}b, it is clear that both the PM-C and SPM-C are nearly
mass independent for a particular set of the parameters.
Interestingly, we see that for PM-R1, no effect appears for
different equations of state and cross sections.

\begin{figure}[!t] \centering
 \vskip 0.5cm
\includegraphics[angle=0,width=10cm]{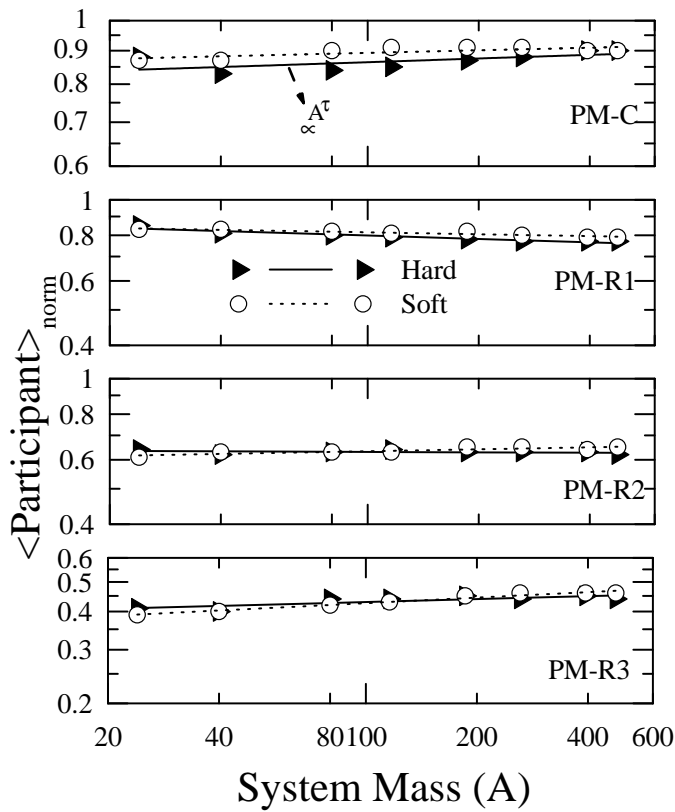}
\vskip -3 cm \caption {Top panel: The participant matter as a
function of system size at their corresponding energy of vanishing
flow. Here we use hard and soft equations of state at $b/b_{max} =
0.4$. The open circles and solid right triangles represent,
respectively, the Soft$^{55}$ and Hard$^{55}$ equations of
state.}\label{fig3}
\end{figure}

In Fig. \ref{fig3}, we display the participant matter as a
function of mass of the system for Hard$^{55}$ (solid right
triangles)
\begin{figure}[!t] \centering
 \vskip 0.5cm
\includegraphics[angle=0,width=10cm]{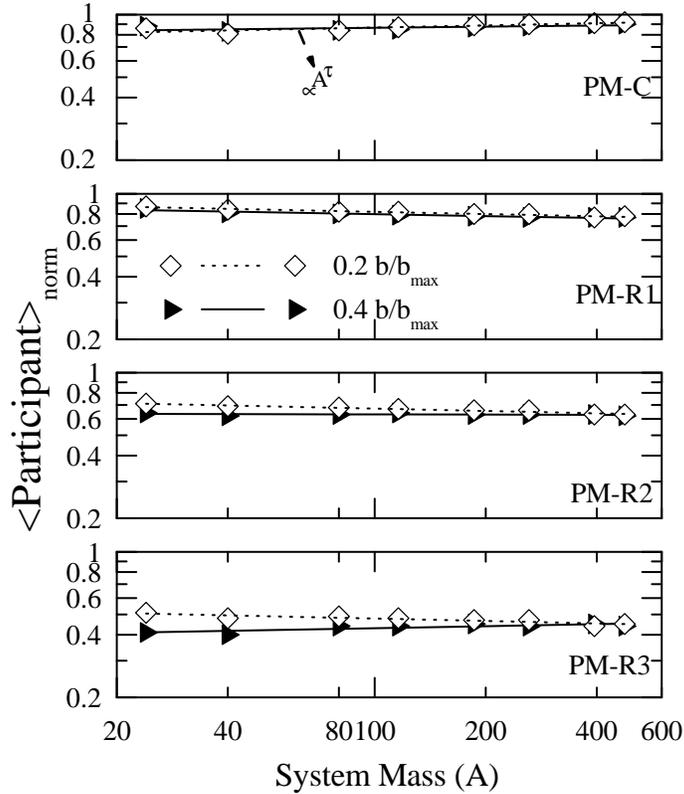}
\vskip -3 cm \caption {Top panel: The participant matter as a
function of system size at their corresponding energy of vanishing
flow. Here we use a hard equation of state along with $\sigma=55$
mb. The open diamonds (solid right triangles) represent the
reactions at $b/b_{max}=0.2$ ($b/b_{max}=0.4$).}\label{fig4}
\end{figure}
and Soft$^{55}$ (open circles) equations of state. Again, lines
are the power law fits ($\propto$$A^{\tau}$). The solid (dotted)
lines are for Hard$^{55}$ (Soft$^{55}$). All types of participant
matter are nearly independent of the equation of state as well as
of the mass of system. Very small dependence on equation of state
is visible for the medium mass systems. This indicates that at
EVF, the sensitivity of participant matter is insignificant
towards the nuclear matter equation of state.

In Fig. \ref{fig4}, we again show the quantities reported in Fig.
\ref{fig1}, for central and semi-central colliding geometries. The
open diamonds are for $b/b_{max}=0.2$ whereas solid left triangles
are for $b/b_{max}=0.4$. Again, almost model and mass independent
behaviour can be seen. This is a very important result since
collective flow is very sensitive to the impact parameter. This
study can be useful in all experimental measurements where impact
parameter variation is not controlled.

From the above discussion, following conclusions are visible.

The participant matter shows almost mass independent behaviour
irrespective of the model parameters for central and semicental
reactions. For a given set of the model parameters, EVF is higher
for lighter colliding nuclei leading to higher density. This
results in frequent nucleon-nucleon collisions resulting in the
mass independent behaviour of participant matter. Note that the
mass independent nature of participant matter is a very important
observation. Especially since participant and spectator matter
serve as an indicator of the role of repulsive and attractive
forces. The contribution of the mean field towards transverse flow
is independent of mass of the system \cite{blat91}. Therefore, one
needs same amount of participant matter to counterbalance
attractive forces.

To understand this further,
\begin{figure}[!t] \centering
 \vskip 0.5cm
\includegraphics[angle=0,width=10cm]{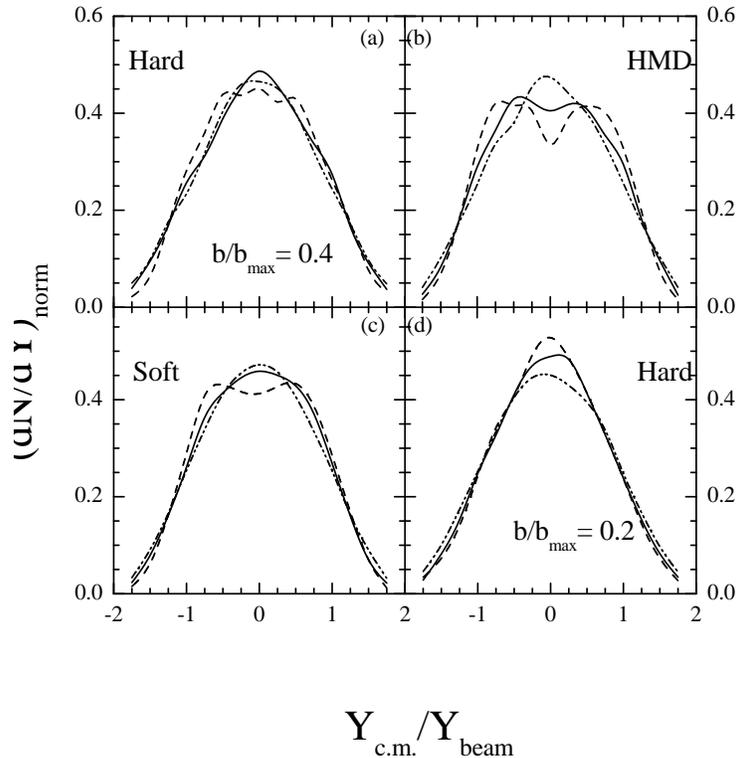}
\vskip -2cm \caption {Top panel: The rapidity distributions for
the reactions of $^{40}$Ca+$^{40}$Ca (dashed line),
$^{93}$Nb+$^{93}$Nb (solid line), and $^{197}$Au+$^{197}$Au
(dash-double-dotted line). The different sub-figures represent the
rapidity distributions for different model
parameters.}\label{fig5}
\end{figure}
we display in Fig. \ref{fig5}, the final state normalized rapidity
distribution. Here we display the reactions of
$^{40}$Ca+$^{40}$Ca, $^{93}$Nb+$^{93}$Nb, and
$^{197}$Au+$^{197}$Au at their corresponding EVF. As already
mentioned, the rapidity distribution is an indicator of the
thermalization achieved in a heavy-ion reaction. From Fig.
\ref{fig5}c (for example), we see that the degree of
thermalization reached is nearly the same irrespective of the mass
of system. This points towards the mass independent nature of
participant matter defined in terms of rapidity distribution cuts.
We also see that the thermalization reached at balance point is
also independent of different cross sections, momentum dependent
interactions, and equation of state. This suggests an
insignificant role of the model ingredients. We have also carried
out similar calculations for the global anisotropy ratio and local
relative momentum. The above mass and model ingredient picture is
also valid in these cases.\\

\section{Summary}

Our present aim was to study the participant-spectator matter over
a wide range of energies of vanishing flow and masses. For this,
we employed different model parameters at central and semi-central
colliding geometries. Remarkably, a nearly mass independent nature
of participant matter was observed at the energy of vanishing
flow. This makes it a very strong alternative candidate to study
the energy of vanishing flow. We have also shown that the
participant matter can act as an indicator to study the degree of
thermalization. The degree of thermalization reached in central
collisions is nearly same for different colliding nuclei at the
energy of vanishing flow.


\begin{thebibliography}{0}
\bibitem{stoc86}H. Stoecker and W. Greiner, Phys. Rep. {\bf 137},
277 (1986).

\bibitem{aich91} J. Aichelin, Phys. Rep. {\bf 202}, 233 (1991);
J. Aichelin, A. Rosenhauer, G. Peilert, H. Stoecker, and W.
Greiner, Phys. Rev. Lett. {\bf 58}, 1926 (1987).

\bibitem{grei88}J. Aichelin, G. Peilert, A. Bohnet, A. Rosenhauer,
H. Stocker, and W. Greiner Phys. Rev. C \textbf{37}, 2451 (1988);
U. Eichmann, J. Reinhardt, and W. Greiner, Phys. Rev. C
\textbf{61}, 064901 (2000); J. Brachmann, S. Soff, A. Dumitru, H.
Stocker, J. A. Maruhn, W. Greiner, L. V. Bravina, and D. H.
Rischke, Phys. Rev. C \textbf{61}, 024909 (2000); S. A. Bass, M.
Hofmann, M. Bleicher, L Bravina, E. Zabrodin, H. Stocker, and W.
Greiner, Phys. Rev. C \textbf{60}, 021901(R) (1999).

\bibitem{hart98} C. Hartnack, R. K. Puri, J. Aichelin, J. Konopka, S. A. Bass,
H. Stoecker, and W. Greiner, Eur. Phys. J {\bf A1}, 151 (1998).

\bibitem{puri94} R. K. Puri, E. Lehmann, A. Faessler, and S. W. Huang,
J. Phys. {\bf G20}, 1817 (1994); R. K. Puri, E. Lehmann, A.
Faessler, and S. W. Huang, Z. Phys. {\bf A351}, 59 (1995).

\bibitem{gupt88} G. F. Bertsch and S. Das Gupta, Phys. Rep. {\bf
160}, 189 (1988).

\bibitem{khoa92548} D. T. Khoa {\it et al.}, Nucl. Phys. {\bf A548}, 102 (1992); D.
T. Khoa {\it et al.}, Nucl. Phys. {\bf A542}, 671 (1992).

\bibitem{ogil90} C. A. Ogilvie {\it et al.}, Phys. Rev. C {\bf 42}, R10 (1990).

\bibitem{krof92} D. Krofcheck {\it et al.}, Phys. Rev. C {\bf 46}, 1416 (1992).

\bibitem{mage0061} D. J. Magestro {\it et al.}, Phys. Rev. C {\bf 61}, 021602(R)
(2000).

\bibitem{mage0062} D. J. Magestro, W. Bauer, and G. D. Westfall,
Phys. Rev. C {\bf 62}, 041603(R) (2000).

\bibitem{cuss02} D. Cussol {\it et al.}, Phys. Rev. C {\bf 65}, 044604 (2002).

\bibitem{west93} G. D. Westfall {\it et al.}, Phys. Rev. Lett. {\bf 71}, 1986
(1993).

\bibitem{sull90} J. P. Sullivan {\it et al.}, Phys. Lett. {\bf B249}, 8 (1990).

\bibitem{ange97} J. C. Angelique {\it et al.}, Nucl. Phys. {\bf
A614}, 261 (1997).

\bibitem{krof91} D. Krofcheck {\it et al.}, Phys. Rev. C {\bf 43}, 350 (1991).

\bibitem{he96} Z. Y. He {\it et al.}, Nucl. Phys. {\bf A598}, 248 (1996).

\bibitem{buta95} A. Buta {\it et al.}, Nucl. Phys. {\bf A584}, 397 (1995).

\bibitem{pak97} R. Pak {\it et al.}, Phys. Rev. Lett. {\bf 78}, 1022 (1997);
R. Pak {\it et al.}, Phys. Rev. C {\bf 54}, 2457 (1996); R. Pak
{\it et al.}, Phys. Rev. C {\bf 53}, R1469 (1996).

\bibitem{zhan90} W. M. Zhang {\it et al.}, Phys. Rev. C {\bf 42}, R491 (1990);
M. D. Partlan {\it et al.}, Phys. Rev. Lett. {\bf 75}, 2100
(1995); P. Crochet {\it et al.}, Nucl. Phys. {\bf A624}, 755
(1997).

\bibitem{li93} B. A. Li, Phys. Rev. C {\bf 48}, 2415 (1993).

\bibitem{mota92} V. de la Mota, F. Sebille, M. Farine, B. Remaud, and P. Schuck,
Phys. Rev. C {\bf 46}, 677 (1992).

\bibitem{xu92} H. M. Xu, Phys. Rev. C {\bf 46}, R389 (1992);
H. M. Xu, Phys. Rev. Lett. {\bf 67}, 2769 (1991).

\bibitem{zhou94a} H. Zhou, Z. Li, Y. Zhuo, and G. Mao, Nucl. Phys. {\bf A580}, 627
(1994).

\bibitem{lehm96} E. Lehmann, A. Faessler, J. Zipprich, R. K. Puri, and
S. W. Huang, Z. Phys. {\bf A355}, 55 (1996).

\bibitem{soff95} S. Soff, S. A. Bass, C. Hartnack, H. Stocker, and
W. Greiner, Phys. Rev. C {\bf 51}, 3320 (1995).

\bibitem{kuma98} S. Kumar, M. K. Sharma, R. K. Puri, K. P. Singh, and I. M. Govil,
Phys. Rev. C {\bf 58}, 3494 (1998); S. Kumar, Ph.D. thesis, Panjab
University, Chandigarh, 1999.

\bibitem{sood04} A. D. Sood and R. K. Puri, Phys. Rev. C {\bf 69}, 054612
(2004); A. D. Sood and R. K. Puri, Phys.  Lett. {\bf B594}, 260
(2004); A. D. Sood and R. K. Puri, Phys. Rev. C {\bf 73}, 067602
(2006).

\bibitem{sood04a} A. D. Sood and R. K. Puri, Phys. Rev. C {\bf 70}, 034611 (2004).

\bibitem{blat91} B. Blattel {\it et al.}, Phys. Rev. C {\bf 43}, 2728 (1991).

\bibitem{pan93} Q. Pan and P. Danielewicz, Phys. Rev. Lett. {\bf 70}, 2062
(1993); J. Zhang, S. Das Gupta, and C. Gale, Phys. Rev. C {\bf
50}, 1617 (1994); J. Singh, S. Kumar, and R. K. Puri, Phys. Rev. C
{\bf 63}, 054603 (2001).

\bibitem{fopi} W. Reisdorf, Nucl. Phys. \textbf{A630}, 15c (1998).

\end{thebibliography}
\end{document}